\documentclass[sigconf,natbib=true,anonymous=false]{acmart}
\usepackage{soul}
\usepackage{bm}
\usepackage[table]{xcolor}
\usepackage{amsmath}
\usepackage{amsfonts}
\usepackage{multirow}
\usepackage{graphicx}
\usepackage{subcaption}
\usepackage{balance}
\usepackage{float}
\usepackage{enumitem}
\usepackage{algorithmic}
\usepackage[ruled, linesnumbered]{algorithm2e}
\usepackage{booktabs} 
\usepackage{makecell} 
\usepackage{multirow}  
\usepackage{booktabs}
\usepackage{siunitx}

\usepackage{algorithm}
\usepackage{algorithmic}
\usepackage{xcolor}

\renewcommand\footnotetextcopyrightpermission[1]{} 
\newcommand{\shortname}{\emph{STEAM}}

\AtBeginDocument{%
  \providecommand\BibTeX{\normalfont B\kern-0.5em{\scshape i\kern-0.25em b}\kern-0.8em\TeX}%
}

\begin{document}

\title{From Atom to Community: Structured and Evolving Agent Memory for User Behavior Modeling}
\author{Yuxin Liao}
\affiliation{
\department{Key Laboratory of Knowledge Engineering with Big Data,}
\institution{Hefei University of Technology}
\city{Hefei}
\country{China}
}
\email{yuxinliao314@gmail.com}


\author{Le Wu}
\authornote{Le Wu is the corresponding author.}
\affiliation{
\department{Key Laboratory of Knowledge Engineering with Big Data,}
\institution{Hefei University of Technology}
\city{Hefei}
\country{China}
}
\email{lewu.ustc@gmail.com}

\author{Min Hou}
\orcid{0000-0002-0524-6806}
\affiliation{%
  \institution{Key Laboratory of Knowledge Engineering with Big Data,}
  \department{Hefei University of Technology}
  \city{Hefei}
  \state{Anhui}
  \country{China}
}
\email{hmhoumin@gmail.com}

\author{Yu Wang}
\orcid{0009-0008-6272-8714}
\affiliation{
    \department{Key Laboratory of Knowledge Engineering with Big Data,}
  \institution{Hefei University of Technology}
  \city{Hefei}
  \country{China}}
\email{wangyu20001162@gmail.com}

\author{Han Wu}
\affiliation{
  \department{Key Laboratory of Knowledge Engineering with Big Data,}
  \institution{Hefei University of Technology}
  \city{Hefei}
  \country{China}}
\email{ustcwuhan@gmail.com}




\author{Meng Wang}
\affiliation{
\department{Key Laboratory of Knowledge Engineering with Big Data,}
\institution{Hefei University of Technology}
\city{Hefei}
\country{China}
}
\email{eric.mengwang@gmail.com}

\begin{abstract}

User behavior modeling lies at the heart of personalized applications such as recommender systems and intelligent assistants. With the rise of Large Language Model (LLM)-based agents, user preference representation has evolved from latent embeddings to semantic memory. While existing memory mechanisms have shown promise in textual dialogue scenarios, modeling non-textual user behaviors remains a significant challenge, as preferences must be inferred from implicit signals like clicks and purchases without ground truth supervision. Current approaches typically rely on a single, unstructured summary to represent user preferences, updated through simple overwriting. However, this design is inherently suboptimal: users exhibit multi-faceted interests that get conflated in a single summary, preferences evolve over time yet naive overwriting causes interest forgetting, and individual interactions are often sparse, necessitating collaborative signals from similar users.
To address this, we present \shortname~(\textit{\textbf{ST}ructured and \textbf{E}volving \textbf{A}gent \textbf{M}emory}), a novel memory framework that reimagines how agent memory is organized and updated for user behavior modeling. \shortname~ decomposes user preferences into fine-grained atomic memory units, each capturing a distinct interest dimension with explicit links to observed behaviors. To exploit collaborative patterns, \shortname~ organizes semantically similar memories across users into communities and generates prototype memories for efficient signal propagation. The framework further incorporates adaptive evolution mechanisms, including consolidation for refining existing memories and formation for capturing emerging interests. Extensive experiments on three real-world recommendation datasets demonstrate that \shortname~ substantially outperforms state-of-the-art baselines in recommendation accuracy, user simulation fidelity, and diversity.
\end{abstract}



\maketitle

\section{Introduction}
\begin{figure}[htbp]
  \centering
  \includegraphics[width=1.0\linewidth]{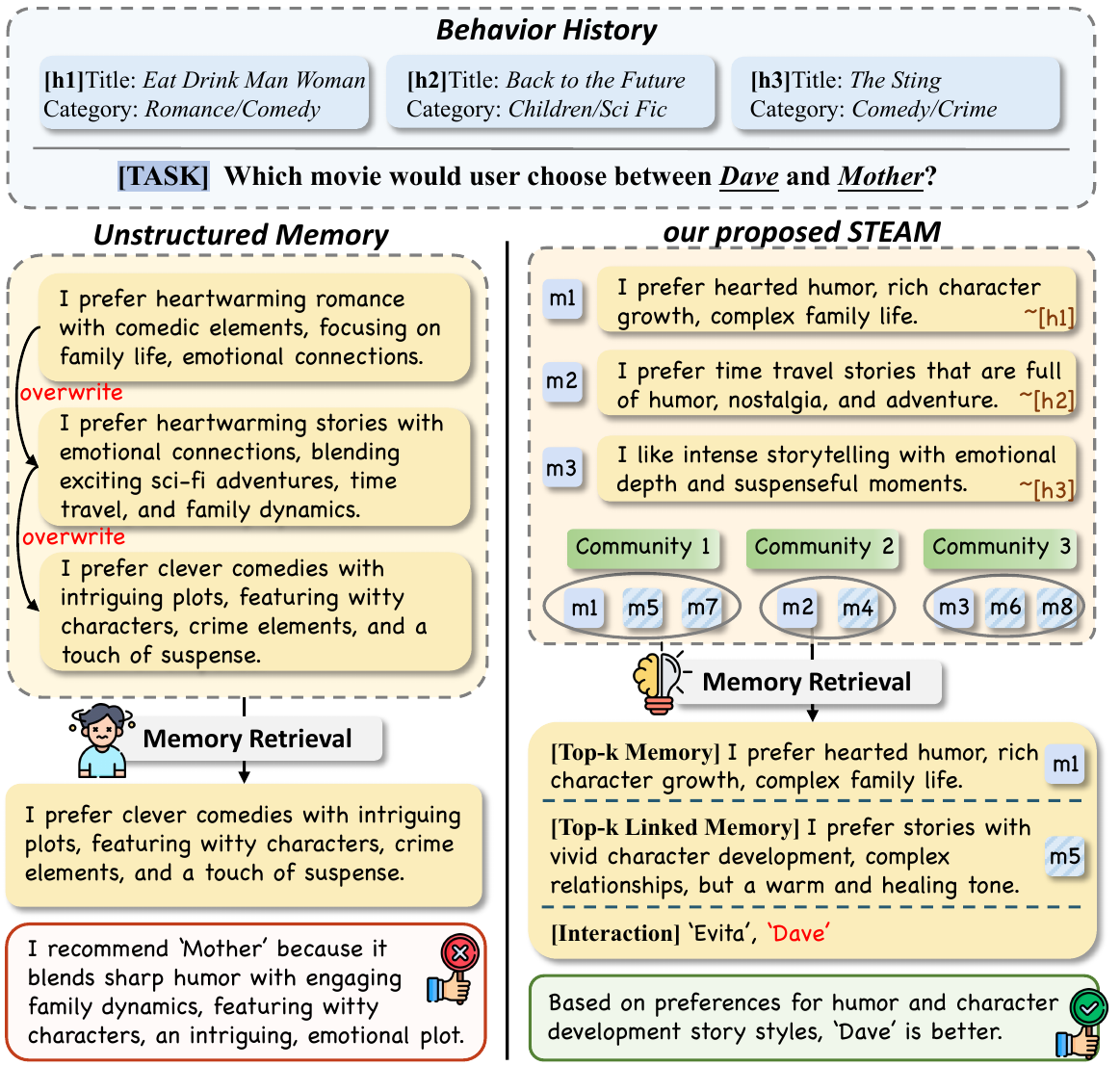}
  \caption{Comparison of unstructured single summary memory and our proposed \shortname~ for user behavior modeling in real recommendation scenario.}
  \label{fig:motivation}
\end{figure}

LLM-based agents~\cite{agentsurvey1,agentsurvey2} have emerged as a powerful paradigm capable of understanding complex contexts and generating human-level intelligent responses.
Recently, there has been a surge of interest in employing LLM-based agents for user modeling~\cite{recagent,agent4usermodeling}, where agents are used to model and mimic user behaviors in user-centric applications such as personal assistants~\cite{pasurvey} and recommender systems~\cite{agentrecsurvey1,agentrecsurvey2,agentrecsurvey3}.

Memory~\cite{memorysurvey1,memorysurvey2,memorysurvey3} is a fundamental component in LLM-based agents for user modeling. It refers to the structures and mechanisms that store, manage, and retrieve user characteristics accumulated from interactions. It enables agents to maintain behavioral consistency and generate personalized responses.
Existing research on memory mechanisms has predominantly focused on textual dialogue scenarios~\cite{dsmart,rmm,memorybank}, such as multi-turn conversations~\cite{locomo} and personal assistants~\cite{memweaver}. However, memory for non-textual user behaviors, such as clicks and purchases, remains relatively underexplored. Unlike textual dialogues, where information is primarily expressed in natural language, user behaviors are driven by implicit user preferences. Given this distinction, the mainstream paradigm involves constructing memory by inferring user preferences from implicit behavioral signals. Memory then guides the prediction of subsequent actions and evolves through self-reflection based on real user feedback. The absence of ground truth regarding user preferences restricts memory to evolving indirectly, posing significant challenges for memory optimization. As shown in the left portion of Fig.~\ref{fig:motivation}, existing methods~\cite{2024agent4rec,2024agentcf,2025agentcf++,2025afl,2024kgla} maintain agent memory merely as a single, unstructured summary of user preferences, updating it through overwriting when new behaviors occur. Such an approach is inherently suboptimal. We argue that user behavior modeling imposes specific challenges on memory mechanisms, which are mainly reflected in the following aspects:
\begin{itemize}[leftmargin=*]
    \item \textbf{Disentangling Multi-Dimensional Preferences.} Users often have multiple interest dimensions simultaneously. As shown in Fig.~\ref{fig:motivation}, a user may primarily enjoy comedy films while also having an interest in sci-fi genres. Each dimension represents a distinct preference characterized by different contextual factors. Unstructured memory that compresses all preferences into a single summary leads LLMs to allocate attention to irrelevant information. This calls for disentangled user modeling, enabling targeted retrieval and updates.

    \item \textbf{Controllable Evolution of Dynamic Interests.} User interests evolve continuously, with new interests emerging and existing preferences shifting. When memory evolves only through naive overwriting, redundancy and semantic conflicts accumulate, leading to degraded memory quality over time. Thus, user behavior modeling calls for controllable memory evolution mechanisms that can consolidate semantically related memories and form new memory units when emerging preferences are observed.

    \item \textbf{Exploiting Fine-Grained Collaborative Signals.} Recommender systems, as a typical application of user behavior modeling, often suffer from interaction sparsity. Collaborative filtering, which leverages preferences from similar users, has proven effective in alleviating this sparsity. However, user agents remain isolated, with each agent maintaining independent memory without cross-user associations. This isolation hinders effective utilization of collaborative information for capturing popularity trends and facilitating serendipitous discoveries.

\end{itemize}

Taken together, these observations indicate that agent memory in user behavior modeling should go beyond textual accumulation. Instead, ideal agent memory is a structured system composed of relatively independent yet interconnected memory units, with explicitly support controlled evolution and relationship building. This principle is consistent with event segmentation theory in cognitive science~\cite{lilun1,lilun2}. 
According to this theory, humans naturally segment continuous experiences into discrete and meaningful events and encode them with temporal and semantic relations.


In this work, we transform the existing single summary memory paradigm into a compositional view where memory consists of fine-grained atomic memory units. Just as atoms constitute matter, these units collectively represent users' diverse interests. Specifically, we propose \shortname~(\textit{\textbf{ST}ructured and \textbf{E}volving \textbf{A}gent \textbf{M}emory}), a self-evolving memory framework for user behavior modeling. As shown in the right portion of Fig.~\ref{fig:motivation}, \shortname~ constructs structured memory by establishing dual correspondences: each memory corresponds to specific user interaction behaviors, and is further associated with memory communities that organize semantically related memories across users, enabling collaborative signal exploitation. Upon this structured foundation, we design mechanisms for community construction and memory evolution. For community construction, \shortname~ captures multi-hop substructures of memory connections to identify collaborative users, and constructs prototype memories to represent each community. This enables multi-order collaborative signal exploration and low-cost propagation without graph convolution. For dynamic preference modeling, \shortname~ consolidates memories reflecting similar interests and forms new memory when emerging preferences are observed. Both self-reflection and collaborative signal propagation are executed asynchronously in batches, ensuring efficiency. We conduct extensive experiments in recommendation scenarios, a representative application of user behavior modeling, to validate the effectiveness of \shortname.

    
    
We summarize the contributions of this work as follows:
\begin{itemize}[leftmargin=*]
    \item We propose a novel structured memory paradigm for user behavior modeling. User memory is composed of fine-grained atomic memory units, enabling disentangled representation that can be retrieved and updated in a targeted manner.
    
    \item We construct memory communities that organize semantically related memories across users, enabling efficient collaborative signal exploration and propagation.
    
    \item Extensive experiments on real-world datasets demonstrate that \shortname~ significantly enhances user behavior modeling.
\end{itemize}

\section{Preliminaries}

\subsection{Problem Formulation}
Suppose we have a user set $\mathcal{U}$ and an item set $\mathcal{I}$, where $|\mathcal{U}|$ and $|\mathcal{I}|$ denote the number of users and items, respectively. For each user $u \in \mathcal{U}$, the historical behavior sequence is represented as $s_t = \{i_1, i_2, \dots, i_t\}$, spanning from timestamp $1$ to $t$. Given a set of candidate items $I_c \subseteq \mathcal{I}$, the objective is to generate a ranked list of recommendations along with explanations based on the understanding of user preferences. In this context, effective user behavior modeling serves as the foundation for accurate next behavior prediction in user-centric applications such as recommender systems.

\subsection{Memory-Centric User Modeling Pipeline}
Following~\cite{2024agentcf,2025afl}, each user $u$ is assigned a dedicated agent equipped with a memory module $M_u$. The memory module maintains and manages the agent's understanding of user preferences through three core operations: \textit{construction} ($\mathcal{F}_M^{con}$), \textit{retrieval} ($\mathcal{F}_M^{ret}$), and \textit{evolution} ($\mathcal{F}_M^{evo}$). During the training phase, agents leverage their memories to simulate user behaviors and update them through self-reflection based on actual behaviors. With the optimized memories, agents generate personalized content during the inference phase.

\subsubsection{Training Phase.}
Taking user $u$ as an example, the training process iteratively invokes the three memory operations as follows.

\textbf{Memory Construction.} At timestamp $t=0$, the memory module is initialized using the user's static profile $p_u$:
\begin{equation}
    M_u \leftarrow \mathcal{F}_M^{con}(p_u).
\label{eq:init}
\end{equation}

\textbf{Memory Retrieval.} At each subsequent timestamp $t$, the agent is tasked with simulating user $u$ to select between two candidates $i_{pos}$ and $i_{neg}$ given the current interaction sequence $s_t$. The agent first retrieves relevant memories by querying with the semantic embeddings $e_{pos}$ and $e_{neg}$ of the candidates. Then it prompts the LLM to generate a response conditioned on the retrieved memory $m_t$ and the candidates' textual information $x_{pos}$ and $x_{neg}$:
\begin{equation}
\begin{aligned}
    &m_t \leftarrow \mathcal{F}_M^{ret}(e_{pos}, e_{neg}), \\
    &a_t = \text{LLM}(m_t, x_{pos}, x_{neg}, \mathcal{Q}),
\end{aligned}
\label{eq:action}
\end{equation}
\noindent where $a_t$ represents the agent's predicted selection and $\mathcal{Q}$ is the task specification prompt.

\textbf{Memory Evolution.} Upon executing an action, the agent compares its simulated action $a_t$ with the real user behavior $i_{t+1}$ to reflect on the decision and derive insights for memory evolution:
\begin{equation}
r_t = \text{LLM}(a_t, i_{t+1}),
\label{eq:reflection}
\end{equation}
where $r_t$ denotes the reflection output generated by the LLM. Based on $r_t$, the agent evolves its memory through $\mathcal{F}_M^{evo}$.

The entire interaction episode can thus be represented as:
\begin{equation}
    \tau = \{(s_1, a_1, r_1), (s_2, a_2, r_2), \dots, (s_t, a_t, r_t)\}.
\end{equation}
Through this training process, the agent continuously adapts to evolving user interests and progressively refines its memory.

\subsubsection{Inference Phase.}
During inference, we formulate the evaluation as a ranking task to assess the fidelity of agent-based user behavior modeling. The agent receives a candidate item set for the user's next interaction. By leveraging the optimized memory module and interaction history, the agent generates a personalized ranking over candidates. We adopt recommendation accuracy and serendipity as evaluation metrics.

\begin{figure*}[t]
  \centering
  \includegraphics[width=\textwidth]{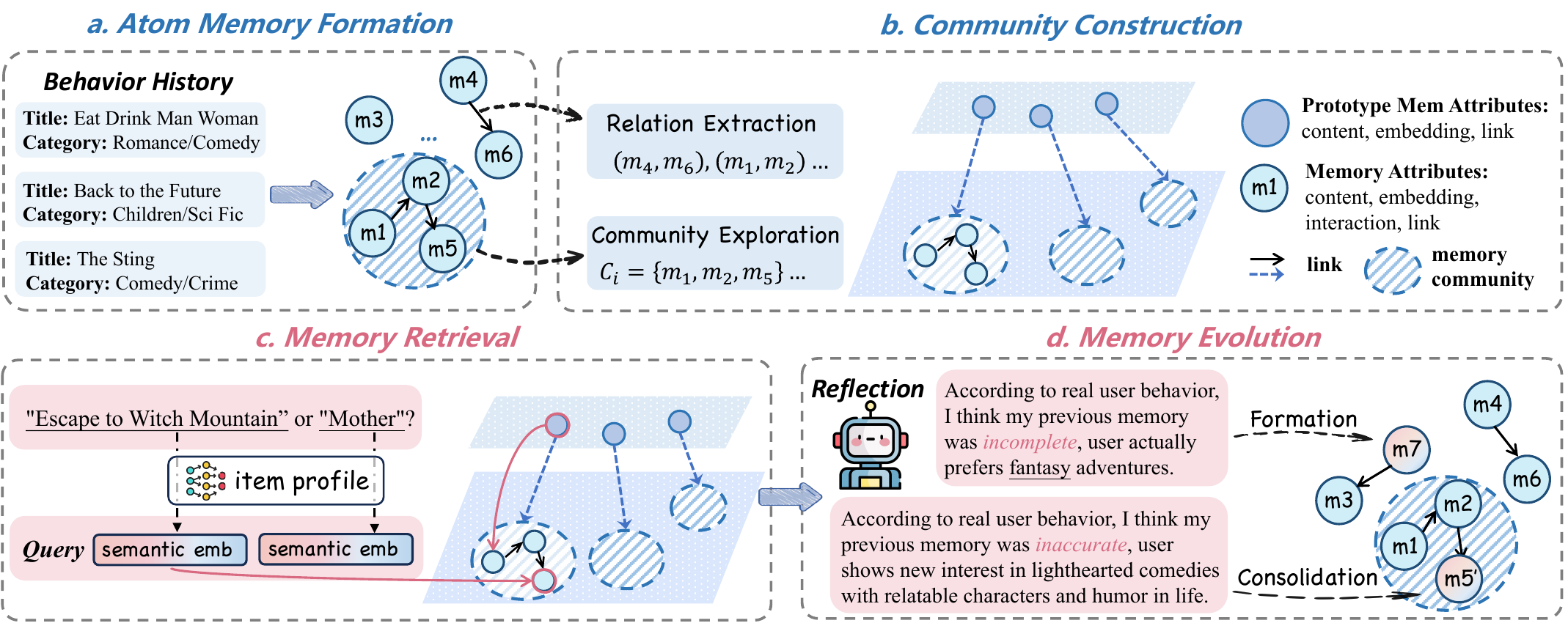}
  \caption{Overview of \shortname. (a) and (b) together illustrate the memory construction process: (a) \textit{Memory Formation} captures user preferences from behavior history. (b) \textit{Community Construction} builds first-order collaborative relationships via relation extraction, explores communities through 2-hop BFS, and summarizes them as prototype memories. (c) \textit{Memory Retrieval} first queries local agent memory, then indexes collaborative memories via corresponding prototypes. (d) \textit{Memory Evolution} reflects on simulated behaviors to consolidate existing memory or form new ones.}
  \label{fig:framework}
\end{figure*}

\section{METHODOLOGY}
\subsection{Overview of \shortname}
As illustrated in Fig.~\ref{fig:framework}, we propose \shortname, a systematic framework for agent-based user behavior modeling. \shortname~ models user memory as a composition of fine-grained atomic memory units, disentangling users' diverse interests. We construct a multi-attribute semantic structure for each memory to establish dual correspondences: each memory corresponds to specific user interaction behaviors, and is further associated with memory communities that organize semantically related memories across users.

Upon this structured foundation, \shortname~ operates through three mechanisms iteratively. \textbf{Memory Formation} captures user preferences from behavior history and constructs new memories when emerging interests are discovered. \textbf{Community Construction} builds first-order collaborative relationships via relation extraction, explores communities through multi-hop substructures, and summarizes them as prototype memories, enabling multi-order collaborative signal exploration and low-cost propagation. \textbf{Memory Evolution} consolidates memories reflecting similar interests or triggers new formation when novel preferences are observed, initiating subsequent community construction. This forms a continuous cycle adapting to evolving user interests.

During \textbf{Memory Retrieval}, \shortname~ operates in a local-to-global manner. The agent first queries relevant memories from its local memory module, then indexes collaborative memories via corresponding prototypes. Within the community, the agent identifies memories from collaborative users and leverages their interaction records for enhanced prediction.

\subsection{Memory Construction}
\subsubsection{Atomic Memory Formation}
\label{sec:AM}
We transform the memory paradigm in user behavior modeling into a set of atomic memory units. Each memory is a structured representation capturing a distinct aspect of user interest, consisting of content $c$, semantic embedding $e$, associated interaction behaviors $I$, and community link $L$:
\begin{equation}
\begin{aligned}
    & m = \{ c, e, I, L \}, \\
    & e = f_{\text{enc}}(c),
    \label{eq:formation}
\end{aligned}
\end{equation}
where $c$ describes the preference in natural language, $e$ is computed by a text encoder $f_{\text{enc}}$, $I$ records interaction behaviors such as item clicks or purchases, and $L$ stores the link to the prototype memory indicating community membership (detailed in Section~\ref{sec:CC}).

At timestamp $t=0$, the agent memory is initialized from user's static profile. During self-reflection, if the agent identifies a new aspect of user interest, a new memory is formed following Eq.(\ref{eq:formation}).

For cross-user collaborative signal discovery, we assign each memory a global index and store its semantic embedding in a shared space. This embedding-only storage reduces overhead while preserving data privacy.

\subsubsection{Community Construction}
\label{sec:CC}
We construct hierarchical structure to propagate collaborative signals through two steps:
\paragraph{\textbf{Relation Extraction.}} 
\shortname~ implements autonomous relation extraction that enables memories to discover semantic neighbors, which we regard as \textit{first-order collaborative relationships}. When a new memory $m_i$ is formed, we leverage its semantic embedding $e_i$ to perform similarity retrieval within the shared embedding space. For each memory $m_j$, we compute the similarity score:
\begin{equation}
    \text{sim}_{i,j} = \frac{e_i \cdot e_j}{\|e_i\| \|e_j\|}.
    \label{eq:sim}
\end{equation}
We then identify the top-$k_1$ most similar memories $\mathcal{N}_{i}$ with similarity scores exceeding a threshold $\tau$:
\begin{equation}
    \mathcal{N}_{i} = \{ m_j \mid \text{sim}_{i,j} \ge \tau,\; \text{rank}(\text{sim}_{i,j}) \le k_1,\; m_j \in \mathcal{M} \},
    \label{eq:top k}
\end{equation}
where $\mathcal{M}$ denotes the set of all memories from other user agents. Based on $\mathcal{N}_{i}$, \shortname~ enables high-order collaborative signal propagation through \textit{Community Exploration}.

\paragraph{\textbf{Community Exploration.}}
\shortname~ constructs a two-level memory architecture where the lower level comprises all memories representing fine-grained preferences, and the upper level consists of prototype memories generalizing coarse-grained preference patterns. 
In practice, we aim to dynamically integrate newly generated user preferences into the memory hierarchy with minimal computational overhead. When a batch of new memories $\{m_i\}_{i=1}^{B}$ is generated, we treat them and their linked neighbors $\mathcal{N}_i$ as nodes, with the semantic links as edges, thereby constructing a graph. 
Unlike traditional graph-based methods, we forgo graph convolution for exploring high-order collaborative signals and instead adopt \textit{2-hop connected subgraph search}. This avoids frequent LLM calls for memory summarization that graph convolution would require in textual scenarios.

Specifically, we start from each memory in the batch and perform Breadth-First Search (BFS) along graph edges within 2 hops:
\begin{equation}
    \mathcal{C}_i = \{ m_j \mid d(m_i, m_j) \leq 2 \},
    \label{eq:bfs}
\end{equation}
where $d(m_i, m_j)$ denotes the shortest path length between $m_i$ and $m_j$. We define the identified connected components as \textit{memory communities}, where memories from different users are linked through shared semantic relationships. Through our structured design, retrieving collaborative memories simultaneously provides access to their interaction attributes, offering direct item-level collaborative signals for recommendation.

To efficiently handle dynamic updates, we adopt an incremental maintenance strategy. When community exploration causes changes in the membership of existing memories, we add all first-order neighbors of the affected memories to the update queue for iterative propagation. This mechanism ensures that memory communities capture cascading effects, where a single new edge may trigger the merging of multiple communities.

For each memory community $\mathcal{C}_k$, we generate a corresponding prototype memory $p_k$ with summarized content $c_k$, semantic embedding $e_k$, and member links $L_k$:
\begin{equation}
    p_k = \{ c_k, e_k, L_k \},
    \label{eq:prototype}
\end{equation}
where $e_k = f_{\text{enc}}(c_k)$ and $L_k = \mathcal{C}_k$. Correspondingly, the link attribute of all member memories is updated to reference their prototype memory:
\begin{equation}
    L_i \leftarrow L_i \cup \{p_k\}, \quad \forall m_i \in \mathcal{C}_k.
    \label{eq:link}
\end{equation}
Through this structure, users with similar interests become interconnected via shared prototype memories, enabling effective utilization of collaborative signals during inference.

\subsection{Memory Retrieval}
\shortname~ operates in a local-to-global manner during memory retrieval. At local level, given a candidate item set $I_c$, the agent computes the semantic embeddings of each candidate and retrieves the top-$k_2$ relevant memories $M_{\text{ret}}$ from its own memory module $M_{u}$:
\begin{equation}
    M_{\text{ret}} = \{ m_j \mid \text{rank}(\text{sim}_{c,j})\le k_2, \; \text{sim}_{c,j} \ge \tau, \; m_j \in M_{u} \},
    \label{eq:ret}
\end{equation}
where $\text{sim}_{c,j}$ is the similarity between candidate item $i_c \in I_c$ and memory $m_j$ computed using Eq.~(\ref{eq:sim}). At global level, the agent queries the link attribute of $M_{\text{ret}}$ to locate the referenced prototype memories and then explores the associated memory communities. Within these communities, the agent identifies collaborative memories and leverages their interaction attributes as direct recommendation signals. Finally, the agent integrates personal preferences with collaborative information as contextual input to the LLM.

\subsection{Memory Evolution}
\label{sec:MC}
After the agent generates recommendations based on retrieved memories, it reflects on the decision by comparing it with the real user behavior. This reflection triggers one of two evolution operations: \textit{consolidation} or \textit{formation}.
If the reflection indicates that existing memories are inaccurate and can be refined through the current interaction, the agent performs \textit{consolidation}. When the newly generated insight exhibits semantic overlap with an existing memory, we leverage the LLM to merge them into unified content $c_{\text{new}}$ and semantic embedding $e_{\text{new}}$, while combining the interaction attributes as $I_{\text{new}}$. The existing memory $m_j$ is then updated while preserving its established link to the prototype memory.
\begin{equation}
    m_j \leftarrow \{ c_{\text{new}}, e_{\text{new}}, I_{\text{new}}, L_j \}.
    \label{eq:merge2}
\end{equation}
Through consolidation, redundant memories are eliminated and preference representations are refined.
If the reflection reveals a new aspect of user interest not captured by existing memories, the agent performs \textit{formation} to construct a new memory, as detailed in Section~\ref{sec:AM}. The complete procedure of \shortname~ is summarized in Algorithm~\ref{alg:hamrec}. During training, each user agent iteratively performs forward prediction and backward reflection to construct and evolve its memory. During inference, agents leverage optimized memories to generate recommendations without backward reflection.

\subsection{Model Discussion}
Compared to the single summary memory mechanism, \shortname~ adopts a hierarchical structure and employs 2-hop BFS to propagate collaborative signals, introducing additional space complexity. However, this design is crucial for obtaining broader and more effective collaborative signals while remaining efficient and controllable. Specifically, 2-hop BFS reduces time complexity by limiting propagation hops, and prototype memory enables a single LLM call for all atomic memory units in each community, whereas summarizing all related memories would require multiple calls with unpredictable quality. Overall, this design achieves a favorable trade-off between space and time efficiency.

\begin{algorithm}[t]
\caption{\shortname: Training Phase}
\label{alg:hamrec}
\begin{algorithmic}[1]
\REQUIRE User set $\mathcal{U}$, interaction sequences $\{s_u \mid u \in \mathcal{U}\}$, $k_1$, $k_2$, threshold $\tau$
\ENSURE Optimized agent memories $\{M_u \mid u \in \mathcal{U}\}$, memory communities $\mathcal{C}$
\STATE \textbf{Initialize:} Assign each user $u \in \mathcal{U}$ an agent with memory $M_u = [m_0]$ encoding the user profile
\STATE \textbf{Initialize:} Memory graph $\mathcal{G}=(\mathcal{V}, \mathcal{E})$, communities $\mathcal{C}=\emptyset$
\FOR{$t = 1$ to $T$}
    \FOR{each agent $u$ with candidate items $I_c$}
        \STATE \textit{// Forward Stage} \label{line:forward-start}
        \STATE Retrieve relevant memories $M_{\text{ret}}$ via Eq.(\ref{eq:ret})
        \STATE Query the link attribute of $M_{\text{ret}}$ to get collaborative memories and generate action $a_t$ via Eq.(\ref{eq:action}) \label{line:forward-end}
        \STATE \textit{// Backward Stage}
        \STATE Generate reflection $r_t$ given real behavior $i_{t+1}$ via Eq.(\ref{eq:reflection})
        \IF{reflection indicates an inaccurate existing memory}
            \STATE Perform memory \textit{consolidation} via Eq.(\ref{eq:merge2})
        \ELSIF{reflection reveals a new interest}
            \STATE Perform memory \textit{formation} via Eq.(\ref{eq:formation}) to create $m_i$
            \STATE Add $m_i$ to $M_u$ and $\mathcal{V}$
            \STATE Retrieve neighbors $\mathcal{N}_i$ via Eq.(\ref{eq:top k}) and add edges to $\mathcal{E}$
            \STATE Discover communities via BFS (Eq.(\ref{eq:bfs})) and update $\mathcal{C}$
            \STATE Generate prototype memory via Eq.(\ref{eq:prototype})
            \STATE Update the link attribute of affected memories via Eq.(\ref{eq:link})
        \ENDIF
    \ENDFOR
\ENDFOR
\end{algorithmic}
\end{algorithm}
\section{EXPERIMENTS}
In this section, we conduct experiments to demonstrate the effectiveness of our proposed \shortname~ and analyze the underlying factors. We focus on the following important research questions:
\begin{itemize}[leftmargin=*]
    \item \textbf{RQ1 (Recommendation Evaluation)}: Does \shortname~ outperform SOTA baselines across multiple datasets?
    \item \textbf{RQ2 (Simulation Evaluation)}: Does \shortname~ achieve better user simulation performance across multiple datasets?
    \item \textbf{RQ3 (Diversity Evaluation)}: Can \shortname~ provide users with diverse recommendations to enhance user experience?
    \item \textbf{RQ4 (Case Study)}: How does \shortname~ model and leverage collaborative signals?
    \item \textbf{RQ5 (In-depth Analysis)}: How do core components, LLM backbones, and hyperparameters affect \shortname?
\end{itemize}
\subsection{Experimental Setup}
\subsubsection{Datasets.} 
To compare methods across different domains, we conduct our experiments on three widely used datasets:  
\begin{itemize}[leftmargin=*]
    \item \textbf{CDs and Clothing}~\cite{amazon}: These are subsets of the Amazon datasets, including purchase records categorized by products. "CDs" covers CDs and vinyl purchase records, while "Clothing" focuses on clothing, shoes, and jewelry.
    \item \textbf{MovieLens-100k}~\cite{movielens}: A movie dataset containing movie ratings along with user profiles and movie genres.
\end{itemize}
For all datasets, we apply 5-core filtering and sort purchase or viewing behaviors chronologically. We apply the leave-one-out method to divide the interaction data into training, validation, and testing sets. Due to expensive API calls and consistent with previous settings~\cite{2024agentcf, 2024kgla, 2025afl}, we sample 100 users along with their interactions from each dataset. The detailed statistics for each dataset are summarized in Table~\ref{tab:dataset}.

\begin{table}[htbp]
\centering
\caption{Statistics of the preprocessed datasets}
\label{tab:dataset}
\begin{tabular}{lrrrr}
\toprule
\textbf{Datasets} & \textbf{\#Users} & \textbf{\#Items} & \textbf{\#Inters.} & \textbf{Sparsity} \\
\midrule
\textbf{CDs (full)}        & 112{,}135  & 73{,}304  & 1441{,}330  & 99.98\% \\
\quad - Sample    & 100       & 687       & 600        & 99.13\% \\
\midrule
\textbf{Clothings (full)} & 443{,}895 & 353{,}709 & 5{,}956{,}426 & 99.99\% \\
\quad - Sample    & 100       & 690       & 600        & 99.13\% \\
\midrule
\textbf{ML-100K (full)}   & 943       & 1{,}341   & 98{,}996   & 92.17\% \\
\quad - Sample    & 100       & 420       & 600        & 98.57\% \\
\bottomrule
\end{tabular}
\end{table}

\begin{table*}[htbp]
\centering
\caption{Performance Comparison of Methods on Datasets. For the performance on sampled datasets, the best performance for each metric is highlighted in bold and the second-best result underlined. The results on the full dataset are used as a reference. The superscript * indicates the Improvement is statistically significant where the p-value is less than 0.05.}
\renewcommand{\arraystretch}{1.1}
\label{tab:main res}
\begin{tabular}{l|c c c|c c c|c c c}
\toprule
\hline
\multirow{2}{*}{Method} & \multicolumn{3}{c|}{CDs} & \multicolumn{3}{c|}{Clothings} & \multicolumn{3}{c}{ML-100K} \\
\cline{2-10} 
 & NDCG@1 & NDCG@5 & NDCG@10 & NDCG@1 & NDCG@5 & NDCG@10 & NDCG@1 & NDCG@5 & NDCG@10 \\
\hline
Pop & 0.1037& 0.2359& 0.3031& 0.0790& 0.1300& 0.2134& 0.0740& 0.1176& 0.1249\\

BM25 & 0.1267& 0.2463& 0.2197& 0.0947& 0.1511& 0.2243& 0.0640& 0.1044& 0.1186\\
\hline
BPR$_{\text{full}}$ & 0.1754& 0.4191& 0.5236& 0.2154& 0.3312& 0.4748& 0.1053& 0.2143& 0.3222\\

SASRec$_{\text{full}}$ & 0.3509& 0.6857& 0.7374& 0.3804& 0.4896& 0.5906&0.1148 &0.6404 & 0.6608\\

BPR$_{\text{sample}}$ & 0.1123& 0.3093& 0.4126& 0.1039& 0.2116& 0.3134& 0.1000& 0.1800& 0.2029\\

SASRec$_{\text{sample}}$ & 0.1935& \underline{0.4044}& 0.4243& \underline{0.2160}& 0.3143& 0.4200& 0.1921& 0.3128& 0.3962\\
\hline

LLMRank & 0.0500 & 0.1777 & 0.2625 & 0.0600 & 0.1889 & 0.2823 & 0.0500 & 0.2001 & 0.2671 \\

InteRecAgent & 0.0600 & 0.2188 & 0.3026 & 0.1600 & 0.3401 & 0.4622 & 0.1039 & 0.1588 & 0.2028 \\

AgentCF & 0.1700 & 0.3915 & 0.5084 & 0.1700 & 0.3560 & 0.5055 & 0.1400 & 0.3253 & \underline{0.4789} \\

AFL & \underline{0.2100} & 0.3908 & \underline{0.5266} & 0.2100 & \underline{0.3897} & \underline{0.5117} & \underline{0.1500} & \underline{0.3303} & 0.4779 \\
\hline
\rowcolor{cyan!10}
\shortname~ (Ours) 
& \textbf{0.3200*} 
& \textbf{0.5307*} 
& \textbf{0.5982*} 
& \textbf{0.2800*} 
& \textbf{0.5316*} 
& \textbf{0.6133*} 
& \textbf{0.1700*} 
& \textbf{0.4052*} 
& \textbf{0.5064*} \\

\textit{Improv.} & \textit{+52.38\%} & \textit{+31.23\%} & \textit{+13.60\%} & \textit{+29.63\%} & \textit{+36.41\%} & \textit{+19.86\%} & \textit{+13.33\%} & \textit{+22.68\%} & \textit{+5.74\%} \\
\bottomrule
\end{tabular}
\end{table*}

\subsubsection{Baselines.}
We compare \shortname~ with the following recommendation methods, grouped by their user modeling approach:
\begin{itemize}[leftmargin=*]
    \item \textbf{\textit{Without User Representation.}} This category includes rule-based recommendation methods. \textbf{Pop} gives recommendations based on item popularity; \textbf{BM25}~\cite{bm25} ranks candidates based on their textual similarity to users' interacted items.
    \item \textbf{\textit{Latent Embedding Methods.}} We include two traditional recommendation methods that utilize embeddings to represent users and items. \textbf{BPR}~\cite{bpr} optimizes the representations with BPR loss; \textbf{SASRec}~\cite{sasrec} uses self-attention to capture sequential patterns in historical records.
    \item \textbf{\textit{LLM-based Semantic Modeling.}} \textbf{LLMRank}~\cite{2024llmrank} transforms user behavior sequences into natural language, leveraging the contextual understanding capabilities of LLMs to capture user preferences. \textbf{InteRecAgent}~\cite{2025interecagent} extracts and updates user profiles by parsing the conversation history with the user, and generates natural language recommendations aligned with user preferences; \textbf{AgentCF}~\cite{2024agentcf} utilizes agent memory to preserve user preferences and collaborative information, and updates memory by simulating interactions between user agents and item agents; \textbf{AFL}~\cite{2025afl} iteratively optimizes the agent's understanding of user preferences by establishing a feedback loop between the user agent and the recommender agent.
\end{itemize}

\subsubsection{Implementation Details.}
Most of the above baselines are evaluated on sampled datasets. For latent embedding methods, we report performance on both full datasets and sampled datasets. We implement data preprocessing and baseline methods based on open-source frameworks~\cite{recbole, 2024llmrank}. We implement LLM-based methods using GPT-3.5-turbo-16k-0613 and report performance of \shortname~ on different LLM backbones. The temperature is set to 0.2, and threshold $\tau$ is fixed at 0.5. For memory retrieval, we utilize the text model ``all-MiniLM-L6-v2'' for semantic representation.

\begin{table*}[t]
\centering
\caption{The user simulation performance of \shortname~ compared with backbone LLM \textit{GPT-3.5-Turbo} and baseline model \textit{AFL}. Bold results indicate the best results and the second-best result underlined.}
\setlength{\tabcolsep}{6pt}
\renewcommand{\arraystretch}{1.1}
\label{tab:user simulation}
\begin{tabular}{c c| ccc| ccc| ccc}
\toprule
\hline
\multirow{2}{*}{1:$k$} & \multirow{2}{*}{Metric}
& \multicolumn{3}{c|}{CDs}
& \multicolumn{3}{c|}{Clothings}
& \multicolumn{3}{c}{ML-100K} \\
\cline{3-5} \cline{6-8} \cline{9-11}
& 
& GPT-3.5-Turbo & AFL & ours
& GPT-3.5-Turbo & AFL & ours
& GPT-3.5-Turbo & AFL & ours \\
\hline
\multirow{3}{*}{1:1}
& Precision & 0.7359 & \underline{0.7460} & \textbf{0.7537} & 0.7008 & \underline{0.7160} & \textbf{0.7247} & 0.6069 & \underline{0.6883} & \textbf{0.7199} \\
& F1 Score  & 0.5669 & \underline{0.5893} & \textbf{0.6700} & 0.5240 & \underline{0.6221} & \textbf{0.6486} & \underline{0.5926} & 0.4837 & \textbf{0.6393 }\\
\hline
\multirow{3}{*}{1:3}
& Precision & 0.4887 & \underline{0.5300} & \textbf{0.5478} & 0.4379 & \underline{0.4820} & \textbf{0.5000} & 0.3411 & \underline{0.3580} & \textbf{0.4750} \\
& F1 Score  & 0.4686 & \underline{0.4800} & \textbf{0.4929} & 0.5431 & \underline{0.5774} & \textbf{0.5918} & 0.4396 & \underline{0.4526} & \textbf{0.5349} \\
\hline
\multirow{3}{*}{1:9}
& Precision & 0.2260 & \underline{0.3679} & \textbf{0.3787} & 0.2057 & \underline{0.2095} & \textbf{0.3333} & 0.1478 & \underline{0.1526} & \textbf{0.2165} \\
& F1 Score  & 0.3169 & \underline{0.3944} & \textbf{0.4194} & 0.3132 & \underline{0.3216} & \textbf{0.4547} & 0.2356 & \underline{0.2372} & \textbf{0.3100} \\
\hline
\bottomrule
\end{tabular}
\end{table*}

\subsection{Recommendation Evaluation (RQ1)}
\subsubsection{Evaluation Settings.}
For the recommendation task, since the quality of user behavior modeling is difficult to measure directly, we evaluate it through the downstream recommendation task, which provides well-defined evaluation metrics. Following previous work~\cite{2024agentcf,2025agentcf++}, we task user agents with ranking candidate items. We treat users' real next interaction as the ground truth, and construct the candidate list by including the target item together with nine randomly sampled items that the user has not interacted with. For InteRecAgent, we adopt its single turn eval setting. For evaluation, we adopt \textit{NDCG@K} as the metric, where $K \in \{1, 5, 10\}$.
\subsubsection{Recommendation Performance}
The overall comparison results are presented in Table \ref{tab:main res}, indicating the generalization capability and effectiveness of our method in practical recommendation scenarios. The following observations are made:
\begin{itemize}[leftmargin=*]
    \item \textbf{\shortname~ Performance.} \shortname~ consistently outperforms all baselines across different datasets. This improvement demonstrates its ability to comprehensively model multi-dimensional user preferences through structured atomic memories, supporting precise retrieval and targeted updates. Additionally, the formation of new atomic memories and memory consolidation effectively track interest evolution. Notably, \shortname~ achieves more significant improvements on sparser datasets, validating its capability to capture broad and meaningful interest communities through propagation of explicit collaborative signals.
    \item \textbf{Effectiveness of Memory Mechanisms.} Methods with explicit memory significantly outperform those without (e.g., LLMRank). This highlights the inherent limitations of directly applying LLMs to capture user behavior patterns, aligning with the original motivation for semantic memory design.
    \item \textbf{LLM-based vs. Traditional Methods.} On sampled datasets, traditional methods based on latent embeddings generally underperform LLM-based semantic memory approaches. The primary reason lies in insufficient training data, where latent user embeddings may suffer from overfitting. Specifically, sampled datasets contain only approximately 0.04\%, 0.01\%, and 0.61\% of complete datasets respectively, relying solely on interaction data from 100 users. By leveraging world knowledge of LLMs, LLM-based methods can encode semantic information beyond user interactions, thereby reducing dependence on dense behavior signals.
\end{itemize}

\subsection{Simulation Evaluation (RQ2)}
\subsubsection{Motivation}
Since recommendation tasks focus solely on users' current interests, we additionally conduct user simulation following~\cite{2024agent4rec,2025afl} for more comprehensive evaluation of user modeling quality. In real-world scenarios, users' item selections are largely driven by a clear understanding of their own preferences, and a personalized agent is expected to maintain long-term consistency in preference representation. We assess how well agents can simulate real users by evaluating their ability to distinguish items aligning with genuine user preferences from those that do not.
\subsubsection{Evaluation Settings.}
Each agent is initialized with a portion of real interactions and presented with 20 items, where the ratio of positive to negative items is set to 1:$m$, $m \in \{1, 3, 9\}$. Positive items are actual interactions excluded from initialization, while negative items are randomly sampled. The agent's response to each item is treated as a binary decision, and we use \textit{Precision} and \textit{F1 Score} to evaluate performance. For simplicity, we compare our method and AFL with the backbone LLM, omitting other baselines. For backbone evaluation, we combine each user's historical item information with 20 candidate items into a prompt and let the LLM output a binary judgment for each item.
\subsubsection{Simulation Performance}
The empirical discrimination results are reported in Table \ref{tab:user simulation}. \shortname~ consistently outperforms both the backbone LLM and baseline AFL, with particular improvements in \textit{F1 score}. This suggests that by decomposing user preferences into structured atomic memory units, user agents can capture distinct preference dimensions, resulting in a deeper understanding of real user interests and thus faithful behavior simulation. In contrast, agent memory with a single summary, such as AFL, may suffer from interest forgetting due to overwriting during dynamic updates, thus failing to maintain preference consistency.

\begin{figure}[htbp]
  \centering
  \includegraphics[width=1.0\linewidth]{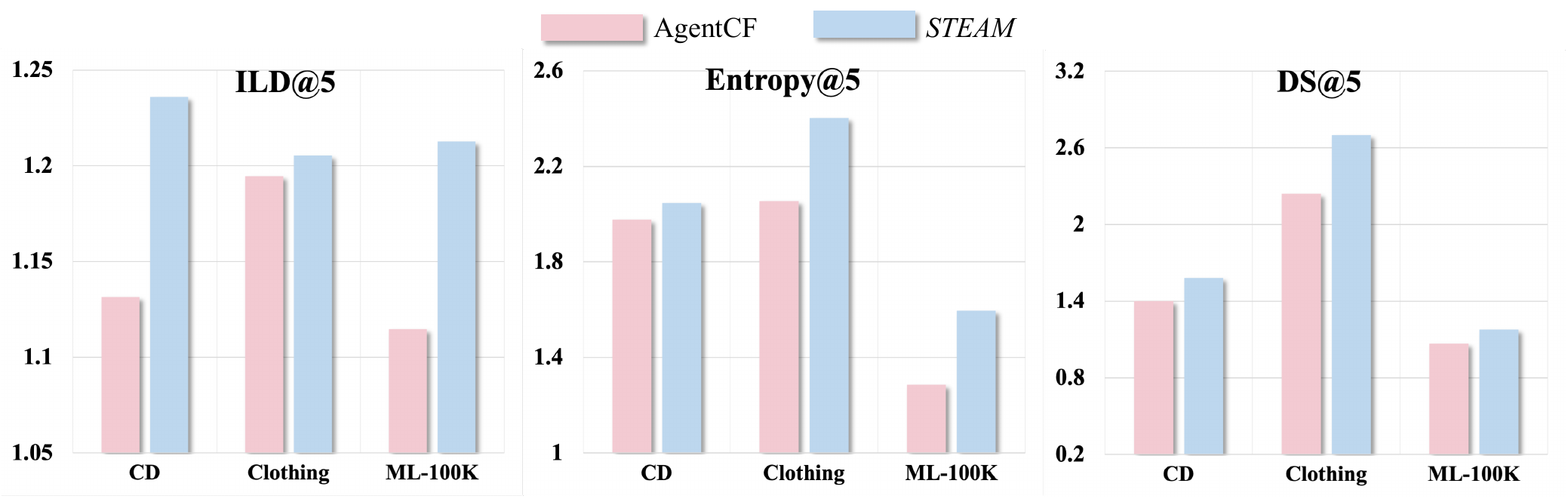}
  \caption{Comparison of recommendation diversity between \textit{AgentCF} and \shortname.}
  \Description{A bar chart comparing recommendation diversity metrics between AgentCF and our proposed method.}
  \label{fig:diverse}
\end{figure}

\subsection{Diversity Evaluation (RQ3)}
\subsubsection{Motivation}
Recommender systems that solely optimize for accuracy tend to over-emphasize dominant interests while neglecting secondary ones, thereby reducing recommendation diversity. With iterative feedback loops, such diversity bias can further lead to information cocoons~\cite{diversity1}. Since \shortname~ explicitly models multiple aspects of user interests, it naturally supports improving recommendation diversity and enhancing user experience.
\subsubsection{Evaluation Settings.}
We focus on individual diversity and adopt three metrics following~\cite{diversity}: \textit{Intra-List Distance (ILD)}, \textit{Entropy}, and \textit{Diversity Score (DS)}, which characterize diversity from the perspectives of item-level dissimilarity, category distribution uniformity, and category coverage breadth, respectively. For the ranked recommendation list $R$ generated by the agent, these metrics are calculated as follows:
\begin{equation}
\small
\begin{gathered}
\mathrm{ILD}(R)=\frac{1}{|R|(|R|-1)}
\sum_{\substack{i,j \in R \\ i \neq j}}
\left\lVert e_i - e_j \right\rVert_2, \\
\mathrm{Entropy}(R)=- \sum_{c \in \mathcal{G}} p(c) \log p(c), \quad p(c) = \frac{1}{|R|} \sum_{i \in R} \mathbb{I}(g_i = c), \\
\mathrm{DS}(R)=\frac{ \left| \{ g_i \mid i \in R \} \right| }{|R|},
\label{eq:diversity}
\end{gathered}
\end{equation}
where $e_i$ denotes the semantic embedding of item $i$, $g_i$ denotes its genre label, and $|R|=10$ is the candidate list size. We report all metrics at cutoff $@5$. For comparison, we select AgentCF as the baseline, since AFL's recommendation lists are influenced by its externally invoked recommendation model, making it less suitable for a fair diversity comparison.
\subsubsection{Diversity Performance}
As illustrated in Fig.~\ref{fig:diverse}, \shortname~ consistently outperforms AgentCF across all diversity metrics on all datasets. AgentCF relies on a single summary to represent user preferences, which inherently biases toward dominant interests, resulting in semantically similar recommendations with narrow category coverage. In contrast, our structured atomic memory units explicitly capture distinct aspects of user interests, enabling the agent to draw upon multiple atoms during recommendation generation and naturally produce diverse yet relevant results. The higher ILD, Entropy, and DS scores confirm that \shortname~ recommends items with greater semantic dissimilarity and broader genre coverage.

\subsection{Case Study (RQ4)}
\begin{figure*}[htbp]
  \centering
  \includegraphics[width=\textwidth]{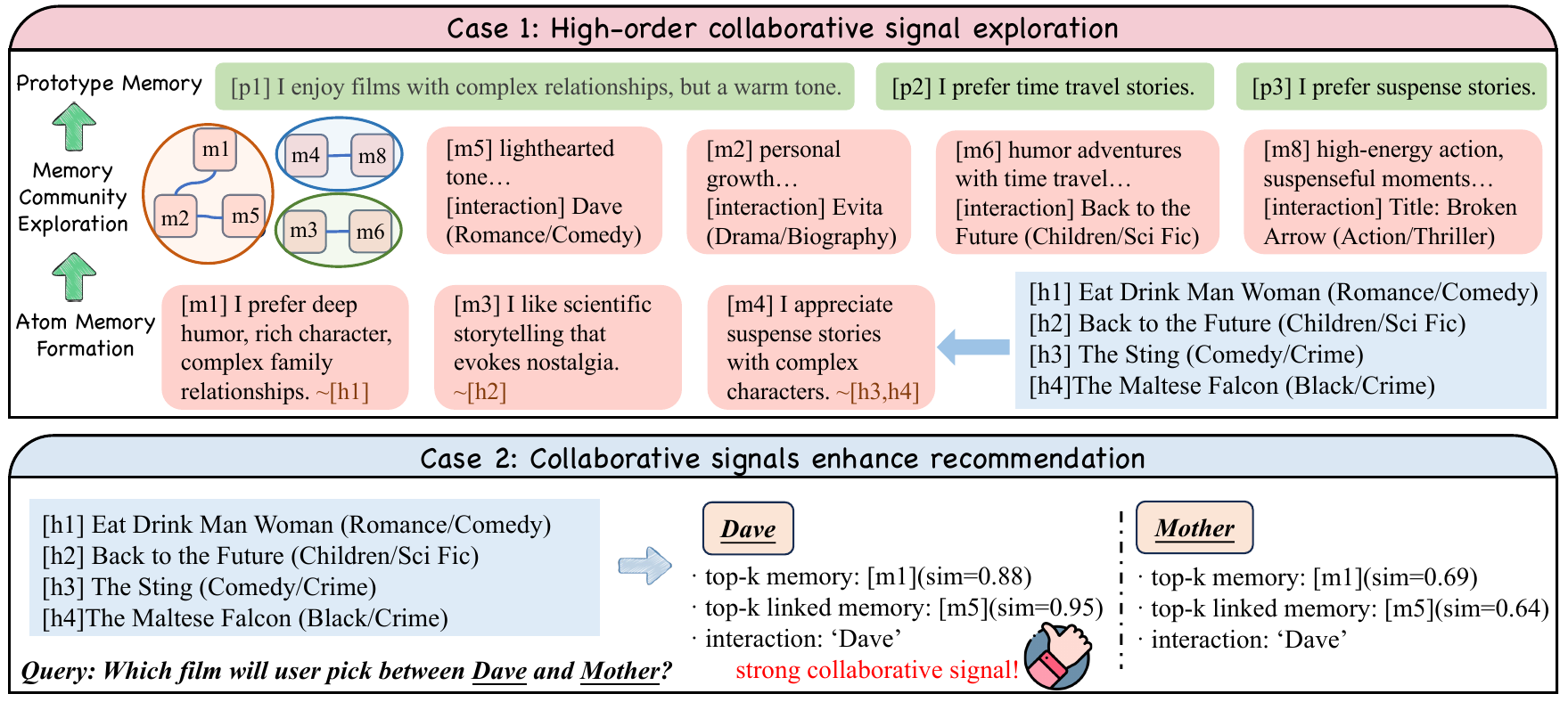}
  \caption{Case studies for collaborative signal exploration and how collaborative signals enhance recommendation.}
  \label{fig:case}
\end{figure*}
To intuitively investigate how \shortname~ models and leverages collaborative signals, we conduct two case studies, as shown in Fig.~\ref{fig:case}. In \textbf{Case 1}, we illustrate the user preference modeling process for a user interaction sequence $\{h_1, h_2, h_3, h_4\}$. During atomic memory formation, both $h_3$ and $h_4$ express the user's preference for suspense stories in the crime genre. Therefore, $m_4$ performs memory consolidation, linking these two interaction behaviors together. During memory community exploration, $m_1$ and $m_2$, as well as $m_2$ and $m_5$, share first-order associations. Thus, through 2-hop BFS, the memory system identifies a community composed of $m_1$, $m_2$, and $m_5$. During prototype memory construction, the formed memory community is summarized, and the atomic memories within the community are linked to the corresponding prototype memory. In \textbf{Case 2}, we demonstrate how the agent leverages the aforementioned memory to enhance recommendation. The agent is tasked with predicting which movie the user will choose between \textit{Dave} and \textit{Mother}. Through memory retrieval, the agent retrieves $m_1$ and further retrieves $m_5$ via the corresponding prototype memory $p_1$, where $m_5$ has the highest similarity to the candidate item \textit{Dave}. By examining the associated interaction attribute, the agent identifies an interaction behavior consistent with the candidate movie. This strong collaborative signal supports the agent in selecting \textit{Dave}.

\subsection{In-depth Analysis (RQ5)}
\subsubsection{Impact of Different LLM Backbones}
To evaluate the impact of different LLMs as backbones, we conduct experiments on three datasets using Claude-haiku-4-5-20251001, Llama-3-70B, and GPT-5.1. As shown in Table \ref{tab: different llms}, \shortname~ achieves consistently strong performance across all LLM backbones and datasets, demonstrating that our structured atomic memory and hierarchical memory architecture are robust and not overly dependent on specific LLM capabilities. This indicates strong generalizability of our memory mechanism design. Furthermore, more advanced LLMs achieve better overall performance, suggesting that stronger reasoning capabilities and richer pretrained knowledge enable the agent to better leverage fine-grained atomic memory units for recommendations.

\begin{table}[ht]
\centering
\caption{Performance comparison (NDCG@10) of \shortname~ with various LLM backbones.}
\label{tab: different llms}
\setlength{\tabcolsep}{8pt} 
\begin{tabular}{c|c|c|c}
\toprule
\hline
\multirow{2}{*}{LLM} & \multicolumn{3}{c}{Dataset} \\ \cline{2-4}
 & CD & Clothing & ML-100K \\ \hline
Claude-Haiku & 0.6247 & 0.6197 & 0.5276 \\  
Llama-3-70B & 0.6283 & 0.6245 & 0.5189 \\ 
GPT-5.1 & 0.6391 & 0.6468 & 0.5418 \\ \hline
\bottomrule
\end{tabular}
\end{table}

\subsubsection{Impact of Different Components of \shortname}
We assess the impact of key components in \shortname~ and provide potential explanations. The following model variants are evaluated:
\begin{itemize}[leftmargin=*]
\item \shortname~ w/o AMU: We replace atomic memory units (Section~\ref{sec:AM}) with a single summary memory adopted by existing methods.
\item \shortname~ w/o MC: We remove memory consolidation in Section~\ref{sec:MC}. When a new atomic memory has semantic overlap with existing memories in the same agent, it will be directly added.
\item \shortname~ w/o CC: We remove community construction in Section~\ref{sec:CC}. This means collaborative information is not leveraged.
\end{itemize}

\begin{table*}[t]
\centering
\caption{Ablation study results on three datasets.}
\label{tab:ablation}
\begin{tabular}{l|ccc|ccc|ccc}
\toprule
\hline
\multirow{2}{*}{Method} & \multicolumn{3}{c|}{CD} & \multicolumn{3}{c|}{Clothing} & \multicolumn{3}{c}{ML-100K} \\
& NDCG@1 & NDCG@5 & NDCG@10 & NDCG@1 & NDCG@5 & NDCG@10 & NDCG@1 & NDCG@5 & NDCG@10 \\
\hline
\shortname~-w/o AMU & 0.2300 & 0.4947 & 0.5765 & 0.2700 & 0.5005 & 0.5782 & 0.1300 & 0.3404 & 0.4796 \\
\shortname~-w/o MC & 0.2600 & 0.5017 & 0.5886 & 0.2300 & 0.4966 & 0.5640 & 0.1400 & 0.3295 & 0.4788 \\
\shortname~-w/o CC & 0.2800 & 0.4992 & 0.5906 & 0.2600 & 0.5023 & 0.5790 & 0.1600 & 0.3691 & 0.5069 \\
\shortname & \textbf{0.3200} & \textbf{0.5307} & \textbf{0.5982} & \textbf{0.2800} & \textbf{0.5316} & \textbf{0.6133} & \textbf{0.1700} & \textbf{0.4052} & \textbf{0.5064} \\
\hline
\bottomrule
\end{tabular}
\end{table*}

Table~\ref{tab:ablation} presents the experimental results on all datasets. We find that atomic memory units (AMU) have the greatest impact on performance; without them, results closely resemble baseline methods with naive memory designs. This confirms that fine-grained preference decomposition is fundamental to precise user modeling. Memory consolidation (MC) is also essential since it maintains memory coherence by eliminating redundancy over long-term simulation. Community construction (CC) also contributes significantly by enabling memory-level collaborative filtering. Its removal leads to notable degradation as user agents lose access to cross-user preference signals.

\subsubsection{Impact of Hyperparameters}
\begin{figure}[htbp]
  \centering
  \includegraphics[width=1.0\linewidth]{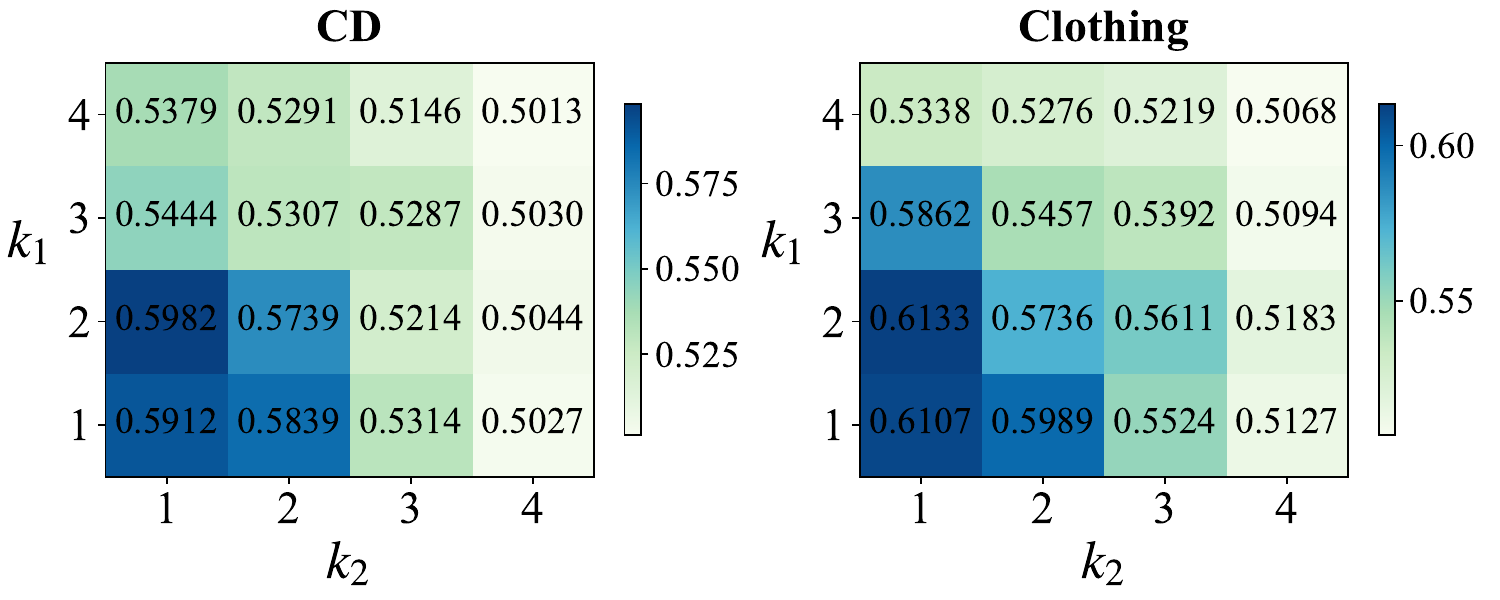}
  \caption{ Performance comparison w.r.t. $k_1$ and $k_2$ of \shortname. The evaluate metric is NDCG@10.}
  \label{fig:param}
\end{figure}

In this part, we analyze the global top-$k_1$ most relevant atomic memories for Relation Extraction and the local top-$k_2$ most relevant atomic memories for Memory Retrieval, corresponding to $k_1$ in Eq.~(\ref{eq:top k}) and $k_2$ in Eq.~(\ref{eq:ret}).
As shown in Figure~\ref{fig:param}, we present results on the CDs and Clothing datasets due to space limitations. The patterns observed across both datasets are generally consistent, with the optimal parameter combination being $k_1=2, k_2=1$. During relation extraction, excessively small values limit discovery of semantically similar memories, thereby constraining the establishment of cross-user links. Conversely, excessively large values introduce noisy connections that blur boundaries of memory communities, degrading the quality of collaborative signals. During memory retrieval, it is essential to retrieve similar memories with high precision. Including preferences with low similarity in the context can distract the LLM.

\section{Related Works}

\subsection{Agent Memory}
Memory has become and will continue to be a core capability of agents built upon LLMs. Through memory mechanisms, agents can accumulate information over interactions and effectively interact with external environments~\cite{memorysurvey1, memorysurvey2, memorysurvey3}. Token-level memory stores factual and personalized information as interpretable token units, and is widely adopted in agent-based personal assistants~\cite{pasurvey} and recommender systems~\cite{agentrecsurvey1}. For factual information, memory retains dialogue content~\cite{rmm} to provide retrievable evidence~\cite{reflexion,buffer}, maintaining contextual consistency in personal assistants~\cite{chatbotmem}. For personalization, memory maintains user profiles~\cite{memorybank} or continuously tracks users' evolving interests~\cite{2024agentcf}, playing an important role in recommender systems where user modeling is central. Despite sharing a similar storage paradigm, memory mechanisms differ substantially in organization and can be broadly categorized into three types. \textit{(i) Flat Memory}~\cite{memorybank,agentkb,mem0,memoryr1} stores memory units without explicitly modeling relationships among them. \textit{(ii) Planar Memory}~\cite{reflexion,memtree,a-mem,ret-mem,m3-agent} extends the flat organization by explicitly capturing relations between memory units, typically using graphs~\cite{a-mem,m3-agent}, trees~\cite{memtree}, or hybrid architectures~\cite{dsmart} to represent their connections, thereby enabling more structured retrieval and reasoning. \textit{(iii) Hierarchical Memory}~\cite{cam,h-mem,g-memory,zep,hipporag,hipporag2} models memory with explicit levels of abstraction, where higher-level representations summarize or abstract lower-level memories. This hierarchical organization allows agents to access information at different levels of granularity, facilitating long-horizon reasoning and complex decision-making.
\subsection{LLM Agents for Recommendation}

LLM-based agents extend LLMs into autonomous systems with mechanisms such as memory, planning, tool use, and self-reflection, enabling iterative perception–reasoning–action loops~\cite{agentsurvey1,agentsurvey2}. This aligns well with the ongoing shift in recommender systems from static preference prediction to dynamic, interactive personalization~\cite{agentrecsurvey1,agentrecsurvey2}. Recent studies can be broadly categorized into three types: recommender agents, conversational agents, and simulator agents. \textit{(i) Recommender Agents}~\cite{recmind, macrec,rah, toolrec, billp} mainly leverage planning and tool use to reason over candidates and produce recommendations; \textit{(ii) Conversational Agents}~\cite{macrs,recllm,2025interecagent,mas,memocrs} focus on eliciting user intent through multi-turn dialogue and providing natural-language recommendations with explanations; \textit{(iii) Simulator Agents}~\cite{2024agent4rec,2024kgla,2025afl,2025agentcf++,2024agentcf,recagent} model users or items as agents with evolving memory to capture user preferences and simulate user behaviors. Recently, researchers have increasingly recognized the importance of memory mechanisms in agents for recommendation. Several approaches~\cite{2024agentcf, 2025agentcf++, memrec} transform isolated memory into collaborative memory. MemRec~\cite{memrec} leverages graph structures to model collaborative relationships. 


\section{Conclusion}
In this paper, we focus on agent memory mechanisms for effective user behavior modeling. To address the key challenges of preference diversity, dynamics, and sparsity, we pioneer the transformation from single summary memory to a composition of fine-grained atomic memory units. Aligned with event segmentation theory in cognitive science, we propose \shortname, a self-evolving memory framework with dual advantages in both effectiveness and efficiency. \shortname~ consolidates and evolves atomic memories to capture preference diversity and dynamics, while organizing them into communities with prototype memories to model multi-order collaborative signals without complex graph convolution. Both self-reflection and collaborative signal propagation are executed asynchronously in batches, ensuring computational efficiency. Extensive experiments on recommendation tasks, a representative application of user behavior modeling, validate the SOTA performance of \shortname. We hope that our proposal will inspire deeper exploration of structured memory mechanisms for building more effective LLM-based agents in user behavior modeling and beyond.

\balance
\bibliographystyle{ACM-Reference-Format}
\bibliography{IL4SR}
\end{document}